Chapter XX

# System Immersion of a Driving Simulator Affects the Oscillatory Brain Activity

*Nikol Figalová [1], Jürgen Pichen [1], Lewis L. Chuang [2], Martin Baumann [1], Olga Pollatos[1]*

[1] *Ulm University, Germany*

[2] *Leibniz Research Centre for Working Environment and Human Factors, Dortmund, Germany*

## ABSTRACT

The technological properties of a system delivering simulation experience are a crucial dimension of immersion. To create a sense of presence and reproduce drivers behaviour as realistically as possible, we need reliable driving simulators that allow drivers to become highly immersed. This study investigates the impact of a system immersion of a driving simulator on the drivers' brain activity while operating a conditionally automated vehicle. Nineteen participants drove approximately 40 minutes while their brain activity was recorded using electroencephalography (EEG). We found a significant effect of the system immersion in the occipital and parietal areas, primarily in the high-Beta bandwidth. No effect was found in the Theta, Alpha, and low-Beta bandwidths. These findings suggest that the system immersion might influence the drivers' physiological arousal, consequently influencing their cognitive and emotional processes.

# INTRODUCTION

Driving simulators allow researchers to study driving behaviour under controlled settings. The need to investigate driver behaviour in a simulated environment is becoming even more pressing with the increasing car automation. The existing automated vehicles cannot be used for real-world studies due to restricted availability or indefensible dangers imposed on the driver or the other road users. However, the simulation does not entail all details of actual driving, and creating a realistic simulation experience proposes a recurring challenge (Hock et al., 2018). Fisher et al. (2011) suggest that participants of driving simulator experiments do not behave the same way they would in their vehicle. In order to reproduce the driver's behaviour as realistically as possible, we need reliable driving simulators which allow participants to get highly immersed in the situation (Michael et al., 2014).

Immersion is a phenomenon experienced by an individual when they are in a state of deep mental involvement. Their cognitive processes cause a shift in their attentional state such that one may experience disassociation from the awareness of the physical world (Agrawal et al., 2020). Nilsson et al. (2016) identified three dimensions of immersion: (a) *system immersion*, i.e. immersion as a property of the technological system delivering the experience and the perceptual response of the user; (b) *challenge immersion*, i.e. immersion as a mental response to challenges; and (c) *narrative immersion*, i.e. immersion as a cognitive response to the narrative characteristics of the experience. The quality of the generated simulation directly moderates the system immersion. A high-fidelity simulator provides higher system immersion, and the overall user's experience feels more realistic, while low-fidelity simulators may evoke unrealistic driving behaviour and, therefore, could produce invalid research outcomes (de Winter et al., 2012).

Measuring immersion using self-report questionnaires can be problematic as the questionnaires rely on participants subjective opinions and requires them to have a fair understanding of what it means to be "immersed" (Jennett et al., 2008). A more objective and reliable way to measure immersion would be beneficial in the research context. With the boom of low-cost and reliable electroencephalography (EEG) headsets, studying brain activity as a proxy of immersion seems viable.

## Neural correlates of immersion

The effect of immersion on brain activity has been previously studied in non-driving contexts. Nacke et al. (2011) reported significant increases in both theta (4-8 Hz) and beta (10-30 Hz) bands with higher immersion during video gameplay. Kruger et al. (2016) investigated the effect of subtitles on an immersion while watching a movie and found that average beta coherence was reduced between the prefrontal and the parietal cortex. This finding suggests greater immersion for the subtitled film viewing experience. Lim et al. (2019) compared concentration and immersion. They suggest that the occipital lobe channel is highly active in the immersion state and report an increase in beta power during an immersion state compared to the rest state and concentration state.

Furthermore, we detected several studies focused on phenomena relevant to immersion. Derbali and Frasson (2010) studied players' motivation in a serious game environment. They observed increased high-Beta power with higher motivation in the area of the left motor cortex (C3). Stenberg (1992) reported an increase in parietal high-Beta bandwidth in conditions with emotional valence. Moreover, he suggests that parietal high-Beta could be interpreted as a measure of alertness or sustained attention. Choi et al. (2015) assessed EEG relative to environmental stress. They found an increase in Alpha and Theta power in non-stressful conditions in parietal (P3) and temporal (T4) regions, respectively, and an increase in high-beta power in the temporal and parietal (P4) region in stressful conditions. Increased motivation, emotional involvement, and stress could indicate higher challenge and narrative immersion in the game, according to the theory of Nilsson et al. (2016).

No studies on the neural correlates of immersion in driving simulator studies were found. Michael et al. (2014) reported higher immersion when using a head-mounted virtual reality device compared to a standard screen; however, only self-report data were collected. Nevertheless, understanding the effect of different simulator settings on brain activity is crucial to ensure that future research is reproducible and consistent. This study aims to explore the effect of system immersion on the drivers' oscillatory brain activity when operating a conditionally automated vehicle (SAE level 3; SAE, 2021) in a driving simulator.

## METHODS

### Experimental design and apparatus

We implemented a between-group design. The high-immersion group (HI) operated a fixed-base driving simulator with a 190 ° field of view, equipped with a realistic car mock-up. The low-immersion group (LI) operated the same simulator, but the side screens providing peripheral visual cues were turned off, and only the front screen was used.

Participants operated a conditionally automated vehicle (level 3; SAE, 2014) through four driving scenarios (approx. 10 minutes each). The order of the scenarios was counterbalanced using the balanced Latin square. The ego vehicle travelled at a set speed of 100 km/h on a rural road with mild oncoming traffic. There was light rain and fog consistently spread throughout the scenarios. Participants simultaneously performed an adapted Visual Search Task (Treisman and Gelade, 1980) on a dashboard screen while the vehicle was operating in an autonomous mode. To minimize the movement artefacts in EEG signal, participants conducted the secondary task using a keyboard placed comfortably in the reach of their hand. Ten auditory take-over requests were issued. These prompted participants to disengage from the secondary task, deactivate the automation, manually overtake a slow vehicle, activate the automation, and return to the secondary task. All participants went through a training session (approx. 15 minutes) before the experiment started.

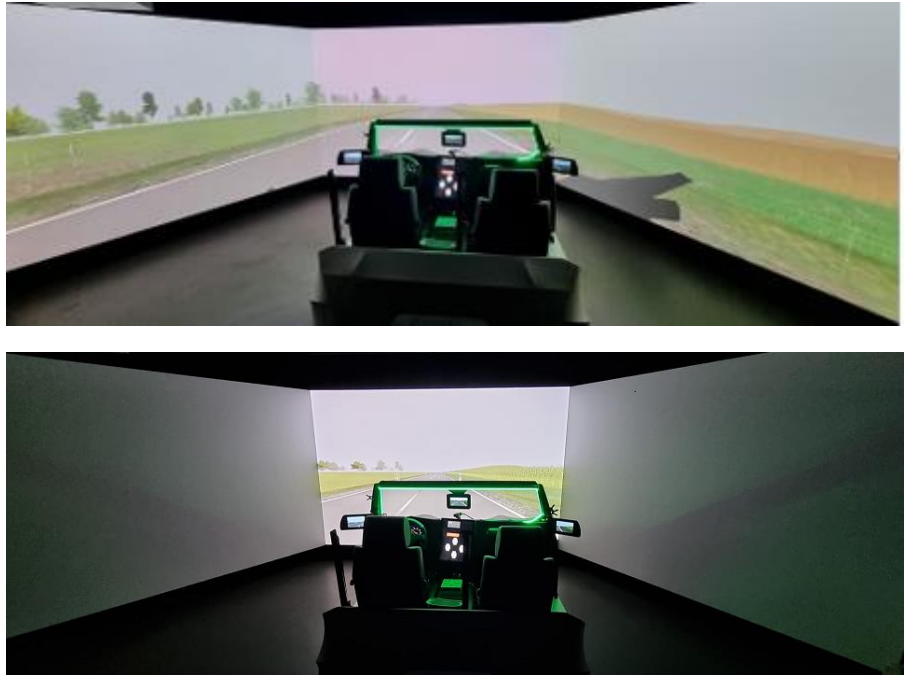

Figure 1: The HI setting (3 screens) and the LI setting (front screen only)

## Participants

We recruited 24 participants; three were excluded from the final analysis due to technical problems, two participants (both from the HI group) were excluded due to simulator sickness. The HI group consisted of nine participants (seven female) with an average age of 22.6 years ($SD = 3.8$). The LI group consisted of ten participants (eight female) with an average age of 23.0 years ($SD = 4.2$). All participants had normal or corrected to normal vision, no neurological or psychiatric disease, and a valid drivers license. Participants were primarily recruited from the Ulm University students and a database of contacts maintained by the dept. Human Factors, Ulm University. All participants were offered either financial compensation or study credits for their participation. All participants provided informed written consent before participating in the experiment. The experimental procedure was performed in accordance with Helsinki's Declaration and approved by the local Ethics Committee of Ulm University.

## EEG data

The EEG was recorded using 32 Ag/AgCl active shielded electrodes placed according to the International 10-20 System with a reference at the FCz position. We focused on the frontal (F3, Fz, F4); parietal (P3, Pz, P4); occipital (O1, Oz, O2); and temporal (T7, T8) regions (see Figure 2). The impedances were kept below 25 kOhm. Horizontal and vertical eye movements were recorded using four electrodes. Data were recorded with a sampling rate of 500 Hz using a LiveAmp amplifier (BrainProducts, Munich, Germany). Noisy electrodes were topographically interpolated, and the signal was re-referenced to a common average reference. A bandpass filter between 0.5 and 40 Hz was applied, as well as a 50 Hz notch filter. The data from the four driving scenarios were segmented into one second long intervals, and independent component analysis (ICA) was used to remove the eye movement artefacts. We used the fast Fourier transformation with Hanning window to calculate the relative spectral power density in the Theta (4-8 Hz), Alpha (8-12 Hz), Beta (12-30 Hz), low-Beta (12.5-16 Hz), and high-Beta (22-30 Hz) range. The power has been normalized as a percentage of the absolute power of the signal in the range from 0.5 to 40 Hz. The pre-processing steps were done using the BrainVision Analyzer 2.1. The Mann-Whitney U test and mixed ANOVAs were conducted using the SPSS 26. The Bayesian statistics were calculated in JASP 0.16.

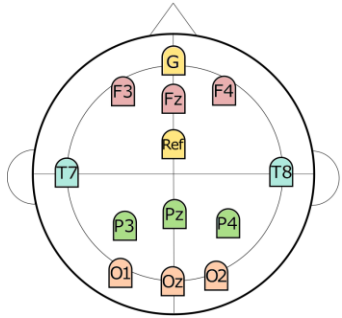

Figure 2: Positions of the electrodes which we used for the analysis

## RESULTS

Firstly, we compared the total relative mean power in the Theta, Alpha, Beta, low-Beta, and high-Beta bandwidths between the LI and the HI group across the four rides. Given the sample size and the characteristics of the data, we used the Mann-Whitney U test. We found no difference between the groups in the Theta, Alpha, and low-Beta bandwidths. There was a significant difference between the groups in the Beta bandwidth at the Cz electrode (LI *Mdn* = 1.05, HI *Mdn* = 1.24, $U = 18.00$, $p = .03$, $\eta^2 = .27$). Moreover, we found a significant difference in high-Beta bandwidth at the Oz electrode (LI *Mdn* = 0.70, HI *Mdn* = 1.20, $U = 19.00$, $p = .04$, $\eta^2 = .25$), the

O2 electrode (LI *Mdn* = 0.58, HI *Mdn* = 1.14, $U$ = 17.00, $p$ = .02, $\eta^2$ = .29), the P3 electrode (LI *Mdn* = 0.65, HI *Mdn* = 0.94, $U$ = 18.00, $p$ = .03, $\eta^2$ = .27), the P8 electrode (LI *Mdn* = 0.63, HI *Mdn* = 1.06, $U$ = 20.00, $p$ = .04, $\eta^2$ = .23), and the Cz electrode (LI *Mdn* = 0.56, HI *Mdn* = 0.86, $U$ = 13.00, $p$ = .01, $\eta^2$ = .38).

To assess the changes over time, we analyzed the differences between the four sequential rides using both frequentist and Bayesian mixed-model ANOVAs. The normality assumption was checked using the Shapiro-Wilk test of normality. As the mixed ANOVA is relatively robust to violations of normality, we proceeded with the analysis when 85% or more of the dependent variables were normally distributed. Due to this requirement, the low-Beta bandwidth was excluded from the analysis as almost half of the dependent variables did not meet the criteria. For the Alpha bandwidth, the two-way mixed ANOVA revealed a significant effect of time in the parietal region ($F(1, 15)$ = 15.74, $p < .001$), temporal region ($F(1,15)$ = 12.56, $p$ = 0.003), and frontal region ($F(1,15)$ = 5.69, $p$ = 0.031). No significant effect was found for group or interaction between time and group. The two-way mixed ANOVA revealed no significant effects in any of the observed regions for the Theta, Beta, and high-Beta bandwidths.

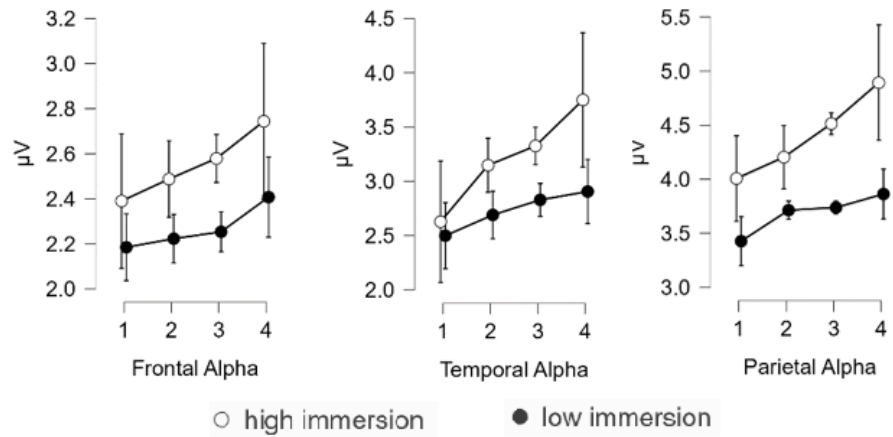

Figure 3: The mean Alpha power over the four consecutive driving scenarios

We calculated the Bayes factor (BF10, Table 1) to understand the results further. The Bayesian approach suggests moderate evidence ($0.33 > BF10 > 0.10$) against the effect of time on frontal and parietal Theta; on occipital and parietal Beta; and occipital high-Beta. Furthermore, we found strong evidence ($BF10 < 0.10$) against the effect of the group on frontal, temporal, and parietal Alpha. Moreover, moderate to strong evidence against the interaction of the two main effects was found on frontal and parietal Theta; occipital, temporal, and parietal Beta; and occipital and parietal high-Beta. Additionally, we found equal support ($BF10 = 1$) for H0 and HA for the effect of time on temporal Theta, frontal, temporal, and parietal Alpha; and equal support for H0 and HA for the effect of Group on parietal Beta and high-Beta.

| Region | Time | Group | Interaction |
|---|---|---|---|
| Theta frontal | **0.144** | 0.803 | **0.118** |
| Theta temporal | 1.000 | 0.610 | 0.729 |
| Theta parietal | **0.144** | 0.3617 | **0.076** |
| Alpha frontal | 1.000 | **0.089** | 0.813 |
| Alpha temporal | 1.000 | **0.004** | 0.846 |
| Alpha parietal | 1.000 | **0.0006** | 0.895 |
| Beta occipital | **0.112** | 0.512 | **0.108** |
| Beta temporal | 0.422 | 0.707 | **0.286** |
| Beta parietal | **0.112** | 1.000 | **0.062** |
| High-Beta occipital | **0.114** | 0.633 | **0.066** |
| High-Beta temporal | 0.854 | 0.653 | 0.591 |
| High-Beta parietal | 0.471 | 1.000 | **0.318** |

Table 1: Results of the Bayesian mixed ANOVA (moderate and strong effects are in **bold**)

## DISCUSSION

In the present experiment, we explored the oscillatory brain activity of a driver while operating a conditionally automated vehicle in a driving simulator. We compared the EEG of drivers operating a highly immersive simulator and drivers operating a low immersive simulator.

We found no significant effect of the system immersion of the simulator on the Theta, Alpha, and low-Beta bandwidths. This finding was further supported by the Bayesian approach, which revealed moderate to strong evidence against the effect of group in the abovementioned bandwidths. This observation does not seem to align with the findings of Nacke et al. (2011), who compared boring and immersive video game experiences. They observed higher activation in the Theta bandwidth when participants experienced the immersive trial compared to the boring trial. However, Nacke et al. did not focus on the system immersion but immersion's challenge and narrative components. They manipulated the architectural complexity of the scenarios. They concluded that the changes in Theta activity originate in the need for episodic and semantic memory activation, path integration, and landmark navigation. None of these was necessary in the present experiment, as we manipulated only the system immersion. Therefore, we believe that our results are not conflicting but rather complementary.

A significant effect of the system immersion of the simulator was found predominantly in the high-Beta bandwidth in the occipital and parietal areas. This result is in line with the findings of Lim et al. (2019), who described an increase in beta power and high activation of the occipital lobe in the immersion state. According to Abhang et al. (2016), high-Beta waves can be associated with stress and high arousal. In the context of the present study, it could suggest that the highly immersive environment leads to higher physiological arousal. As cognition is connected to arousal (Choi et al., 2015; Stenberg, 1992), it would be desirable to study further how the system immersion affects the driver's cognitive process and psychological presence. Moreover, high-Beta waves were associated with emotion (Stenberg, 1992). This could play a significant role in driving simulator experiments, e.g., for user experience assessment or risk perception in simulated environments.

When comparing the four sequential rides, we found an increase in alpha power as the effect of time in both groups. This can be attributed to the time-on-task effect (Figalová et al., 2021; Zhao et al., 2012). However, no effect of group or interaction was found. Based on this observation, we assume that the different setting of the simulator does not induce progressive changes in brain activity over time.

Several limitations can be found in this study. First, the sample size is relatively small, which was also suggested by the results of the Bayesian tests. Collecting a larger sample size would benefit the robustness of the results. Moreover, it would be interesting to compare the self-reported immersion using a scale; however, this was not possible in the present study. In future studies, we also recommend studying the effect of system immersion on simulator sickness, given that only participants from the HI group showed symptoms of simulator sickness. We also recommend combining both the frequentist and Bayesian approach, which seems to be a good way to understand the results comprehensively. Finally, it would be interesting to assess the effect of immersion on the cognitive and emotional processes of the driver.

## CONCLUSIONS

The system immersion of a driving simulator affects the oscillatory brain activity of the driver. We found empirical evidence for changes in the high-Beta bandwidth in the parietal and occipital areas and Beta bandwidth at the Cz electrode. No effect of the system immersion was observed in the Theta, Alpha, and low-Beta bandwidths. The change in the psychophysiological activation suggests that the system immersion modulates the arousal of drivers and could therefore affect their cognitive and emotional processes, perceived stress, and presence.

## ACKNOWLEDGMENTS

This project was funded by the European Union's Horizon 2020 research and innovation program under the Marie Skłodowska-Curie grant agreement 860410.